\documentclass[%
 reprint,
 superscriptaddress,
 showpacs,preprintnumbers,
 amsmath,amssymb,
 aps,
 prl,
 floatfix,
 a4paper
]{revtex4-2}

\usepackage[utf8]{inputenc}

\usepackage{lmodern}
\usepackage{graphicx}% Include figure files
\usepackage{dcolumn}% Align table columns on decimal point
\usepackage{bm}% bold math
\usepackage{isotope,xspace}
\usepackage[squaren,thinspace,mediumqspace,Gray]{SIunits}

\usepackage{xcolor}
\definecolor{medium-blue}{rgb}{0,0,0.5}
\usepackage[colorlinks=true,final,unicode]{hyperref}
\hypersetup{
	pdftitle={Limit on the Fierz interference term b from a measurement of the beta asymmetry in neutron decay},
	pdfauthor={H. Saul, C. Roick, H. Abele, H. Mest, M. Klopf, A. Petoukhov, T. Soldner, X. Wang, D. Werder, and B. Märkisch},
    colorlinks, linkcolor={medium-blue},
    citecolor={medium-blue}, urlcolor={medium-blue}
}

\newcommand{\perkeo}{\textsc{Perkeo~III}\xspace}

\newcommand{\lif}{\isotope[6]{LiF}\xspace}
\newcommand{\sn}{\isotope[113]{Sn}\xspace}
\newcommand{\bi}{\isotope[207]{Bi}\xspace}
\newcommand{\ce}{\isotope[139]{Ce}\xspace}

\newcommand{\cs}{\isotope[137]{Cs}\xspace}

\newcommand{\tum}{Physik-Department ENE, Technische Universität München, James-Franck Straße~1, 85748 Garching, Germany}

\newcommand{\ati}{Technische Universität Wien, Atominstitut, Stadionallee~2, 1020 Wien, Austria}
\newcommand{\hd}{Physikalisches Institut, Universität Heidelberg, Im~Neuenheimer~Feld~226, 69120 Heidelberg, Germany}
\newcommand{\ill}{Institut Laue-Langevin, 71 avenue des Martyrs, CS 20156, 38042 Grenoble Cedex 9, France}
\graphicspath{{figures/}{../figures/}}

\begin{document}

%\preprint{2019-06-26}

\title{Limit on the Fierz Interference Term $b$ from a Measurement of \\the Beta Asymmetry in Neutron Decay}
%\thanks{}%

\author{H.~Saul}
\affiliation{\tum}
\author{C.~Roick}
\affiliation{\tum}

\author{H.~Abele}
\affiliation{\tum}%
\affiliation{\ati}%
\affiliation{\hd}%
\author{H.~Mest}
\affiliation{\hd}%
\author{M. Klopf}
\affiliation{\ati}%
\author{A.K.~Petukhov}
\affiliation{\ill}%
\author{T.~Soldner}
\affiliation{\ill}%
\author{X.~Wang}
\affiliation{\tum}%
\affiliation{\ati}%
\author{D.~Werder}
\affiliation{\hd}%

\author{B.~Märkisch}
\email{maerkisch@ph.tum.de}
\affiliation{\tum}
\affiliation{\hd}

\date{\today}% It is always \today, today,

\begin{abstract}
  In the standard model of particle physics, the weak interaction is described by vector and axial-vector couplings only. Non-zero scalar or tensor interactions would imply an additional contribution to the differential decay rate of the neutron, the Fierz interference term. We derive a limit on this hypothetical term from a measurement using spin-polarized neutrons. This method is statistically less sensitive than the determination from the spectral shape but features much cleaner systematics. We obtain a limit of $b = 0.017(21)$ at $68.27\,\%$ C.L., improving the previous best limit from neutron decay by a factor of four.
%\begin{description}
%\end{description}
\end{abstract}

% see https://publishing.aip.org/publishing/pacs/pacs-2010-regular-edition
% PACS, the Physics and Astronomy Classification Scheme.

\maketitle

%\tableofcontents

%\section{\label{sec:intro}Introduction}

Precision measurements of $\beta$-decays play an important role in understanding the nature of weak interaction and the standard model of particle physics. Neutron beta decay provides a set of observables including the lifetime $\tau_n$, decay spectra and angular correlations. These are used to determine couplings and the Cabibbo-Kobayashi-Maskawa matrix element $V_{ud}$ within the standard model, as well as to perform searches for novel scalar and tensor couplings.
For a beam of polarized neutrons the differential decay rate is described by \cite{Jackson57}:
\begin{multline} \label{eqn:intro:jackson}
\mathrm{d}\Gamma_n\left(E_e,\Omega_e,\Omega_{\nu},\left<\bm{\sigma}_n\right>\right)\mathrm{d}E_e\mathrm{d}\Omega_e\mathrm{d}\Omega_{\nu} = \\
\frac{G_\mathrm{F}^2\left|V_{ud}\right|^2 }{32\pi^5}\rho\left(E_e\right)
(1 + 3 \lambda^2) \left\{1 + a\frac{\bm{p}_e\cdot\bm{p}_{\nu}}{E_eE_{\nu}} + b\frac{m_e}{E_e}\right. + \\
\left.\frac{\left<\bm{\sigma}_n\right>}{\sigma_n}\left[A\frac{\bm{p}_e}{E_e} + B\frac{\bm{p}_{\nu}}{E_{\nu}} + D\frac{\bm{p}_e\times\bm{p}_{\nu}}{E_eE_{\nu}}  \right]\right\},
\end{multline}
with momenta of the electron $p_e$ and the neutrino $p_\nu$, their energies $E_e$ and $E_\nu$, the neutron spin $\bm{\sigma}_n$, the phase space density $\rho(E_e)$ and the decay parameters $a$, $b$, $A$, $B$, and $D$. Within the standard model, the correlation coefficients $a$, $A$ and $B$ depend solely on the ratio of axial-vector and vector coupling constants, $\lambda = g_\mathrm{A}/g_\mathrm{V}$, and the T-violating parameter $D$ is related to the phase of $\lambda$. The beta asymmetry parameter $A$ quantifies the parity-violating beta asymmetry and is most sensitive to~$|\lambda |$. Neglecting order 1\% corrections and assuming $\lambda$ real, the asymmetry $A$ is given by \cite{Wilkinson82}
\begin{equation}
A_0=-2\frac{\lambda(\lambda+1)}{1+3\lambda^2}.
\end{equation}
$A$ has been measured with high accuracy in \cite{Mund13,Brown18,Maerkisch19}.

The shape of the electron spectrum given by the phase space density is independent of whether the weak interaction is generated by vector and axial-vector couplings ($\mathrm{V}-\mathrm{A}$) or scalar and tensor couplings ($\mathrm{S}-\mathrm{T}$) \cite{LeeYang56}.
However, in case of hypothetical scalar or tensor couplings in addition to the $\mathrm{V}-\mathrm{A}$ interaction of the standard model, the beta spectrum would be modified by the Fierz interference term $b$:
\begin{equation} \label{eqn:intro:fierz}
  b \simeq \frac{g_\mathrm{S} g_\mathrm{V} + 3 g_\mathrm{A} g_\mathrm{T}}{g_\mathrm{V}^2 + g_\mathrm{S}^2 + 3\left(g_\mathrm{A}^2 + g_\mathrm{T}^2\right)} \simeq 2\frac{g_\mathrm{S} + 3\lambda g_\mathrm{T}}{1 + 3\lambda^2},
\end{equation}
with coupling constants $g_\mathrm{V,A,S,T}$.
Such a contribution would change the overall decay probability of unpolarized neutrons according to Eq.~\eqref{eqn:intro:jackson}
and also change the measured values $\tilde{X}$ of many of the correlation coefficients $X$, see also Refs.~\cite{Gonzalez-Alonso16,Gonzalez-Alonso18}:
\begin{equation} \label{eqn:theory:bsm:fierz:betaasym}
  \tilde{X} = X\cdot\left(1+\left<b\frac{m_e}{E_e}\right>\right)^{-1}.
\end{equation}
Combining several measured correlation coefficients allows to simultaneously determine $\lambda$ and place limits on scalar and tensor couplings. Alternatively, limits on left-handed tensor interactions can also be obtained by combining the measured beta asymmetry parameter with the stringent limits on scalar interactions from superallowed beta decays \cite{Hardy15}, see Refs.~\cite{Pattie13,Maerkisch19}. For recent reviews, surveys, and limits on couplings beyond the standard model, see Refs.~\cite{Gonzalez-Alonso18,Chang18,Hayen19,Holstein14,Wauters14,Cirigliano13,Bhattacharya12,
Vos15,Dubbers11,Abele08}.
A large Fierz term of $O(10^{-2})$ would be required by the proposed dark decay to explain discrepancies in neutron lifetime measurements  using different methods \cite{Ivanov19}.

In order to put a \emph{direct} constraint on the Fierz interference term $b$ from a single measurement, either the spectral shape of the beta spectrum from unpolarized neutrons or the asymmetry spectrum corresponding to one of the correlation coefficients need to be analyzed.
A first analysis of the unpolarized beta spectrum has been published in \cite{Hickerson17}. Currently, this type of analysis is clearly limited by detector systematics. However, the same principle will be utilized with upcoming measurements to be performed by Nab \cite{Pocanic09,Baessler14,Fry19}, NoMoS \cite{Wang13a,Moser19}, and PERC \cite{perc08,perc12,Wang19}.

In this Letter, we present for the first time an analysis of the experimental beta asymmetry spectrum to derive a limit on the Fierz interference term. While this method provides a smaller statistical sensitivity reduced by about an order of magnitude compared to a limit from the beta spectrum directly \cite{Glueck95a}, detector and background systematics are strongly suppressed. This leads to an improved limit on the Fierz term derived from the largest available dataset in neutron beta decay obtained by the instrument \perkeo of $6 \times 10^8$ decay events. The result is purely limited by statistics.

%\section{\label{sec:perkeo}The spectrometer \textsc{Perkeo~III}}

The analysis presented herein extends the results of \cite{Maerkisch19} to a correlated analysis in $A$ and $b$. For a description of the experimental setup see \cite{Maerkisch09, Maerkisch14b, Maerkisch19}, which include a schematic of the spectrometer. 

\perkeo was installed at the PF1B cold neutron beam facility at the Institut Laue-Langevin, which offers a neutron capture flux density of $\Phi=\unit{2\times 10^{10}}{s^{-1}cm^{-2}}$ \cite{Abele06}. The neutron beam is limited in wavelength to \unit{5}{\angstrom} with about \unit{10}{\%} FWHM by means of a Dornier velocity selector. A bender type supermirror polarizer \cite{Soldner02} is used to select neutrons with a certain spin direction which can be reversed with an adiabatic fast passage spin flipper \cite{Bazhenov93}. After the flipper the neutron beam passes a collimation system consisting of 5 \lif apertures which limit the cross section and divergence of the neutron beam. A rotating disc chopper is mounted directly in front of the spectrometer to create a pulsed neutron beam \cite{Maerkisch09}. During the measurement two different chopper frequencies ($94$ and $\unit{83}{Hz}$) have been used.

The core component of the spectrometer is an $\unit{8}{\meter}$ long, normal conducting magnet system which surrounds the vacuum vessel. The central part of the spectrometer features a nearly homogeneous magnetic field with a maximum of $B_{\textrm{max}} = \unit{152.7}{mT}$ and has a length of $\unit{2.7}{\meter}$. This part acts as the central decay volume in which the charged decay particles from neutron decay are collected by the magnetic field. The charged decay particles are separated from the neutron beam by curved magnetic field sections and are guided to the two detectors. The remaining neutron beam is dumped into a \isotope[10]{B}$_{4}$C beam stop.

The electron detectors consist of Bicron BC-400 plastic scintillators, each read out on two sides by a total of 6 fine-mesh photomultipliers of type Hamamatsu R5504 / R5924. The photomultipliers are connected to the scintillator via acrylic light-guides. The scintillators have a thickness of $\unit{5}{mm}$ which allows full energy deposition of electrons up to an energy of $\unit{1.0}{MeV}$ at normal incidence and minimum sensitivity to $\gamma$-rays. The active area of the detectors is $43 \times 45\ \text{cm}^2$.

In order to monitor and calibrate the detectors, multiple electron conversion sources (\ce, \sn, \cs, \bi) are installed inside the decay volume of \perkeo. These provide five calibration peaks covering an energy range from $\unit{100}{keV}$ to $\unit{1.0}{MeV}$. The sources are deposited on ultra thin carbon foils with a thickness of about $\unit{100}{nm}$. They are mounted on a mechanical apparatus in the central volume allowing to move the sources in two dimensions to measure the detector responses and to map out their uniformity.

%\section{\label{sec:fierz}The Fierz Interference Term}

The data used in this analysis is a subset of the data used in Ref.~\cite{Maerkisch19}. It consists of four datasets each representing one chopper frequency and the detector that triggered first. About $\unit{20}{\%}$ of the data have been excluded due to missing corresponding measurements of the detector uniformity. These data are needed to calculate potentially substantial systematic corrections to this correlated analysis with a potentially large impact on the analysis of $b$. In total about $4.8\times 10^8$ decay events are included in the analysis.

Neglecting quadratic contributions of scalar and tensor interactions to the beta asymmetry parameter $A$, the basic expression that is used in this analysis is the experimental asymmetry spectrum
\begin{equation} \label{eqn:fierz:betaasym}
  A_\mathrm{exp}(E_e) = \frac{N^{\uparrow}(E_e) - N^{\downarrow}(E_e)}{N^{\uparrow}(E_e) + N^{\downarrow}(E_e)} 
   = \frac{v(E_e) A(\lambda) P_n M}{2 c \left(1+b\frac{m_e}{E_e}\right)},
\end{equation}
with $\lambda$ and $b$ as free parameters and with the electron velocity $v(E_e)$, the average neutron beam polarization $\mathbf{P}_n = \langle\bm{\sigma}_n\rangle/\sigma_n$ and the correction $M$ to account for the magnetic mirror effect due to the shape of the magnetic field, see below. The arrows $\uparrow\downarrow$ indicate the neutron spin direction. The factor $1/2$ stems from the integration of the angular cosine distribution.
The full fit function contains theoretical corrections to account for proton recoil, weak magnetism and radiative corrections \cite{Wilkinson82, Ivanov13}. These are functions of $\lambda$ and the electron energy $E_e$. In order to facilitate comparison and use in new physics searches, we provide results in terms of $A(\lambda)$ instead of $\lambda$.

The factors $P_n$ and $M$ represent the dominant experimental corrections to the measurement of the beta asymmetry and have been analyzed separately in order to ensure a \textit{blinded} analysis \cite{Maerkisch19}. They scale the amplitude of the experimental beta asymmetry and thus the result for $A$, whereas $b$ is only affected due to the corresponding change in the theoretical corrections and correlations between the parameters.

\begin{figure*}[t]
\centering
\includegraphics{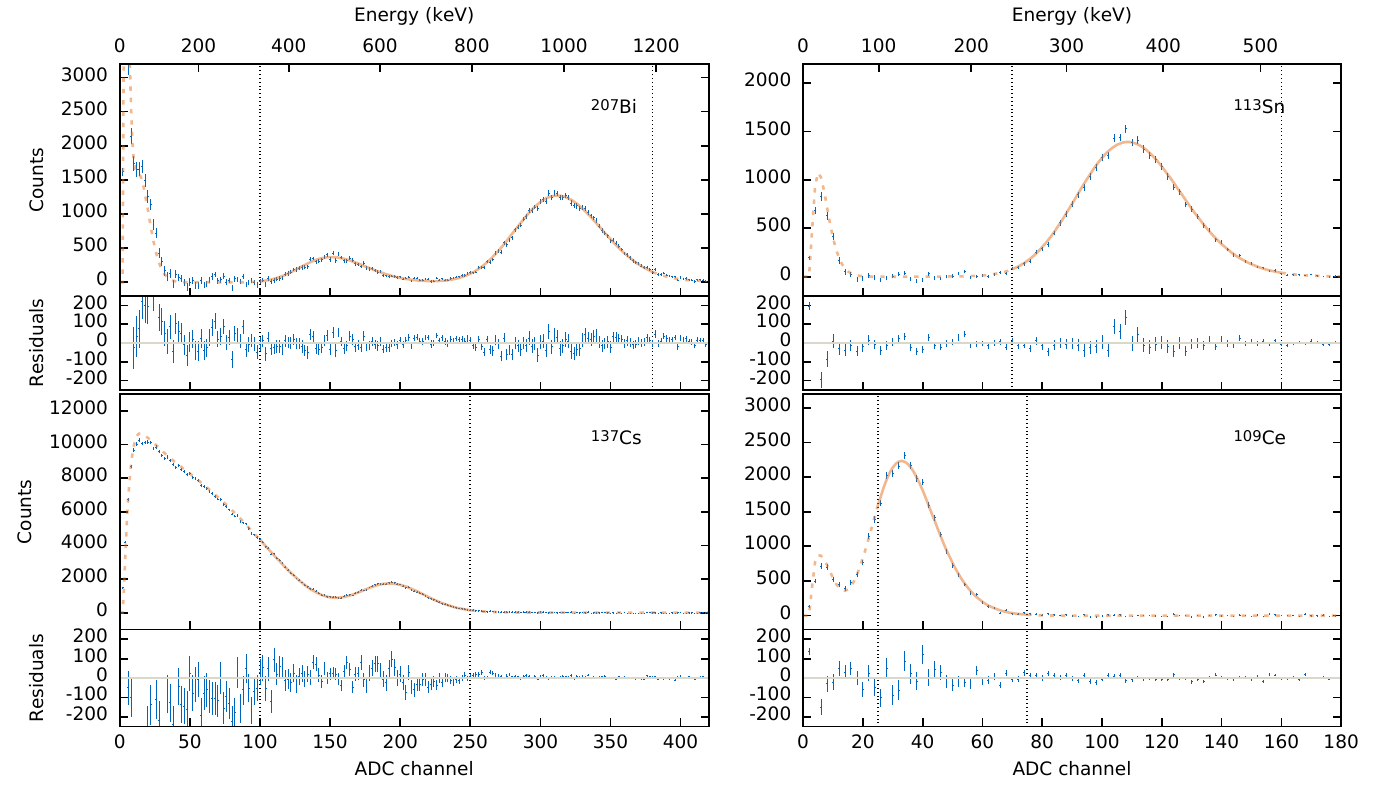}
\caption{Example of measured calibration spectra and the simultaneous fit performed to extract the free parameters of the detector response model. The dashed vertical lines indicate the fit regions. The spectra are practically free of $\gamma$-background from the sources. The discrepancies at low energy outside the fit range mostly stem from uncertainties in the exact shape of the trigger function. We only show every other data-point to improve the presentation.}
\label{fig:spectral:calibration:fit}
\end{figure*}

On top of the underlying fit function a semi-analytical model of the detector response is applied to account for relevant effects. These include non-linearity effects in the scintillator and electronics, broadening of the signal due to photoelectron conversion inside the photomultipliers, the photomultiplier gain process and electronic noise in the data acquisition system, as well as the mapping of the signal amplitude to ADC channels.

The free parameters of the detector response model are determined using a simultaneous fit to measured calibration spectra. The quality of the fits can be seen in Fig.~\ref{fig:spectral:calibration:fit} for one out of 114 calibration sets obtained within the 60 days of measurement. The plastic scintillators of \textsc{Perkeo~III} have a low sensitivity for $\gamma$ radiation and the solid angle for $\gamma$ radiation from the sources onto the detectors is negligibly small.

%\subsection{\label{sec:fierz:systematics}Systematic effects}

Leading systematic effects are related to the detector response: the non-linear detector response function, temperature stability, uniformity and electron backscattering.
The resulting corrections and uncertainties are uncorrelated and small compared to the statistical uncertainties. This allows for separate treatment of the individual systematic effects.
The corresponding uncertainties are extracted as a covariance matrix which is diagonal with respect to the dominant systematic shift observed in $A$ and $b$. This allows to also account for the correlation induced by systematic effects.
The different relevant systematic effects are discussed below; the respective corrections and uncertainties are summarized in Table~\ref{tab:budget}.

\begin{table}[t]
\centering
\begin{ruledtabular}
\begin{tabular}{l c c}
  \textbf{Effect on} & \multicolumn{2}{c}{\textbf{Correction}} \\
\textbf{\boldmath$A$ and \emph{b}}     & \textbf{on \boldmath$A$} &\textbf{on \emph{b}} \\
                                & $\left(10^{-4}\right)$                      & $\left(10^{-3}\right)$  \\[0.15em]\hline\\[-0.9em]
% {-30.8, 0.1, 0.2, -17.3, 9.9, 0, 1.6, 2., 0}
% {-11.4, 0., 0.1, -6.4, 3.7, 0., 0.6, 0.7, 0.}
%{3.7, 0.5, 0.9, 2.5, 4.6, 2.4, 2.7, 4., 1.1, 8.5}
%{1.4, 0.2, 0.3, 0.9, 1.7, 0.9, 1., 1.5, 0.4, 3.2}
Neutron beam				&				& \\
\quad Polarization and		& $-11.4(0.9)$\textsuperscript{a}	& $0.5(0.6)$\textsuperscript{a} \\
\quad Spinflip efficiency	&   	 	& \\
Background					&&\\
\quad Time variation		& \phantom{00}$0.02(0.12)$ & \phantom{0}$0.04(0.24)$\\
\quad Chopper		        & \phantom{00}$0.09(0.22)$ & \phantom{0}$0.5(0.4)$\\
Electrons					&&\\
\quad Magnetic mirror effect	& \phantom{0}$-6.4(0.6)$\textsuperscript{a} & \phantom{0-}$1.0(0.4)$\textsuperscript{a}\\
\quad Undetected backscattering & \phantom{*-0}$3.7(1.1)$	& $-6.4(1.9)$\\
%\quad Lost backscatter energy 	& $0$	 	& $1.4$\\
Electron detector 			&&\\
%\quad Deadtime 			& $(5)$\textsuperscript{*}& $0.35$\\
\quad Temporal stability 	& \phantom{-00.0}$(0.6)$\textsuperscript{a}	& \phantom{-0.0}$(1.2)$\textsuperscript{a}\\
\quad Non-uniformity		& \phantom{-0}$0.6(0.7)$ & $-1.1(1.3)$\\
\quad Non-linearity		& \phantom{-0}$0.8(1.0)$ & $-0.8(1.6)$\\
%\quad Calibration (input data)	&    			& $1$\\
Theory			 			&&\\
\quad Radiative corrections	& \phantom{0}$-1.0(0.3)$\textsuperscript{a} & \phantom{*-}$3.5(0.5)$\textsuperscript{a} \
\\[0.15em]\hline\\[-0.9em]
Total systematics	        & $-12.8(2.1)$ 	        & $-6.1(3.2)$\\
Statistical uncertainty		& \phantom{0.00}$(14.4)$& \phantom{0.00}$(20.4)$ \
\\[0.15em]\hline\\[-0.9em]
\textbf{Total}                  & \phantom{0.00}\boldmath$(14.6)$	& \phantom{0.00}\boldmath$(20.6)$\\
\end{tabular}
\end{ruledtabular}
\caption{\label{tab:budget}
Summary of corrections and uncertainties (in parentheses) to $A$~and~$b$. In contrast to Ref.~\cite{Maerkisch19}, we provide the magnitude of the corrections, not fractional ones. Corrections which contribute only on the $10^{-5}$-level are omitted. One of the fit parameters is actually $\lambda$, not $A$, but we list corrections on $A$ for comparability with earlier measurements.\\
\textsuperscript{a}Already included in the fit shown in Fig.~\ref{fig:fierz:fit}: measured by the data acquisition system or included in the fit function.}
\end{table}

\paragraph{Non-linearity:}
The response of the detectors is not linear with respect to the energy of the detected electrons. One contribution to the non-linearity is caused by saturation effects in the light generation inside the scintillator which is empircally described by the Birks model \cite{Birks51} using the quenching parameter $k_B$. Recent measurements for similar scintillator materials yield \mbox{$k_B = 100\text{--}\unit{150}{nm/keV}$}~\cite{Aberle11} which we confirmed for our scintillator by independent test measurements using Compton recoil electrons from $\gamma$ radiation \cite{saul18}.
Another contribution to the non-linearity arises from bandwidth limitations in the electronics used. Although studied in offline measurements, the determination of the exact contribution of the electronics to the detector non-linearity would require a full reproduction of the original detector setup.
Instead, the analysis is done with three different models to describe the overall non-linearity of the detector response and with and without the calibration point at $\unit{1}{MeV}$. The free parameters of these models are obtained from fits to spectra from calibration sources. A pure Birks model to describe the combined non-linearity of scintillator and electronics leads to $k_B \simeq \unit{400}{nm/keV}$ dependent on the dataset. The maximum difference between the results of the three methods is taken as systematic uncertainty.

\paragraph{Stability:}
The gain of the detector varies over time mainly due to fluctuations in the temperature. This effect can be as large as $\unit{2}{\%}$ between day and night. The detector gain has been monitored hourly using a single calibration source and the signal amplitudes have been renormalized in the data reduction process by interpolating between these hourly measurements.  With this method, the detector gain is effectively stabilized on the $10^{-3}$-level. However, the remaining fluctuations still contribute one of the dominant systematic uncertainties related to the detector response.

\paragraph{Detector Uniformity:}
The detector response varies spatially due to the variation in photon transport efficiency inside the scintillator. This leads to a relative deviation from uniformity by $<\pm\unit{2.5}{\%}$ over the area covered by decay electrons. The spatial response of the detectors was mapped weekly using a single calibration source at different positions. The difference in detector coverage between calibration measurements and measurements of the beta asymmetry leads to a small change in the effective signal response for the two types of measurements. This includes the dependence of the magnetic point-spread of electrons on the beta asymmetry itself \cite{Dubbers14, Dubbers15}. In order to account for these differences, the corresponding spectral corrections are calculated using photon transport simulations performed with \textsc{geant4} \cite{Geant4a} which are matched to the measured spatial detector response. The consistent characterization of the detector without resorting to beta decay data is a major improvement of this measurement.

\paragraph{Undetected Backscattering:}
The symmetric design of \perkeo with two detectors connected by the magnetic field enables reconstruction of the full electron energy even if electrons are backscattered from the scintillator, as both detectors are always read out simultaneously. This backscattering occurs with a probability of about $\unit{11}{\%}$ \cite{Roick19} and is suppressed by reflections at the magnetic field to $\unit{6}{\%}$ of events where the electron reaches the opposite detector (see also \cite{Abele97}). Due to the finite trigger threshold with an efficiency of 50\% at channel $\approx 5.5$ (corresponding to $\approx \unit{30}{keV}$), the energy deposited in the first detector in rare cases is not sufficient to trigger the electronics and the wrong emission direction and energy are assigned to the event. This changes the asymmetry and the shape of the measured asymmetries $A_\mathrm{exp}(E_e)$ for both detectors. An energy-dependent correction for undetected backscattering is included in the analysis, which is calculated based on backscattering simulations performed with \textsc{geant4} \cite{Geant4a} and verified against measured data. Details on undetected backscattering in \perkeo can be found in \cite{Roick19}.

\begin{figure}[t]
  \centering
  \includegraphics[width=0.45\textwidth]{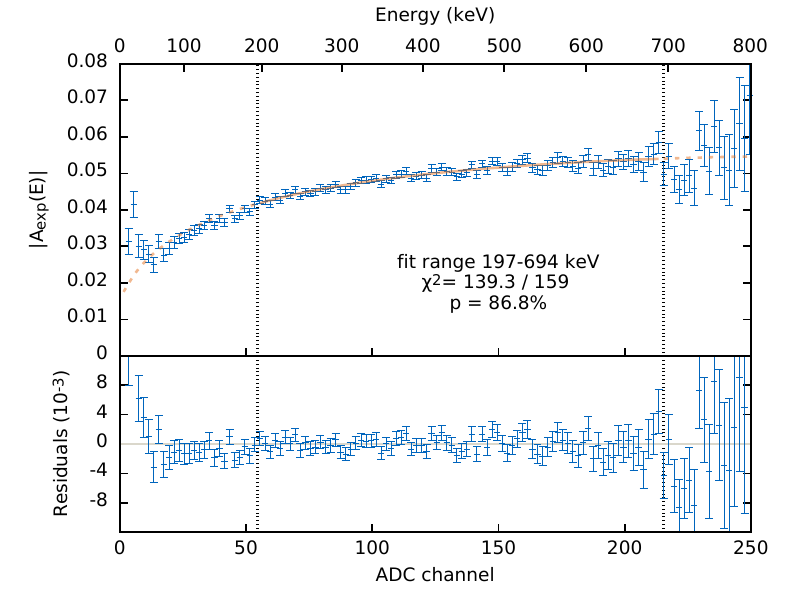}
  \caption{Fit to the data at the chopper frequency of $\unit{83}{Hz}$ measured by the downstream detector. The fit includes theoretical corrections as well as the full model of the detector response. Corrections for detector non-uniformity and undetected backscattering are not included. The data is rebinned by a factor of two for presentation. The fit range is indicated by the dashed vertical lines.}
  \label{fig:fierz:fit}
\end{figure}

\paragraph{Background:}
The pulsed beam allows to select events according to their neutron time-of-flight from the chopper within the spectrometer. The time window is chosen such that the neutron pulse is fully contained in the central decay volume, without any contact to material, and all decay electrons are guided to one of the detectors. After the neutron pulse is absorbed by the beamstop, another time window is used to extract the background. This way signal and background are measured during every chopper cycle with the chopper being closed. In addition to the electron detectors, NaI $\gamma$-detectors have been used to monitor the time dependence of the ambient background and that created by the chopper disc. Both effects have been found to be on the $10^{-5}$ / $10^{-4}$-level for $A$ / $b$, respectively. The effect of the background subtraction can be seen in Fig.~2 of Ref.~\cite{Maerkisch19}.
In Table~\ref{tab:budget} we separately list contributions by the (non-)uniformity of the chopper disc and potential variation in time within the background time window.

%\section{\label{sec:result}Result}

\begin{figure}[t]
  \centering
  \includegraphics[]{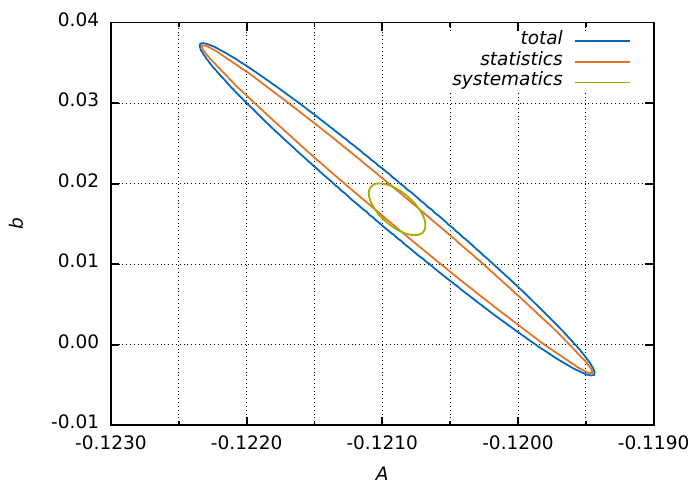}
  \caption{68.27\% confidence region for the beta asymmetry parameter $A$ and Fierz interference parameter $b$ and the contributions by statistics and systematics.}
  \label{fig:fierz:result}
\end{figure}

Fig.~\ref{fig:fierz:fit} shows a fit to one of the four datasets. As was the case already with the single parameter fits in Ref.~\cite{Maerkisch19}, the fit quality is excellent for all individual datasets. The energy range of the fit was extended to $197\text{--}\unit{694}{keV}$ compared to \cite{Maerkisch19} to optimize the total error of this two-parameter fit ($\lambda$ and $b$). The results of all datasets are consistent with statistical fluctuations. We obtain a combined result of
\begin{align}
  A &= -0.1209(14)_{\mathrm{stat}}(2)_{\mathrm{sys}} = -0.1209(15), \nonumber \\
  b &= 0.017(20)_{\mathrm{stat}}(3)_{\mathrm{sys}} = 0.017(21),
    \label{eqn:final-result} \\ 
  \rho_{A,b} &=-0.985, \nonumber
\end{align}
with one sigma errors $(\Delta \chi^2=1)$ and where $\rho_{A,b}$ is the off-diagonal element of the correlation matrix. The fit result for $\lambda$ corresponding to $A$ is $\lambda = -1.2792(60)$. 
This result includes the systematic corrections which are summarized in Table~\ref{tab:budget} and the propagated statistical uncertainties of the free parameters of the detector response model as obtained from the calibration analysis. However, these contributions on the $10^{-5}$ and $10^{-4}$ level for $A$ and $b$, respectively, are negligible. 
Fig.~\ref{fig:fierz:result} compares the error ellipses resulting from statistical and systematic uncertainties, where statistical uncertainties are dominating.

Assuming the standard model with $b \equiv 0$, we obtain from our correlated result in Eq.~\eqref{eqn:final-result}:
\begin{equation}
	\ A = - 0.11972(25), \quad \lambda = -1.27607(68),
\end{equation}
in agreement with our analysis within the standard model in Ref.~\cite{Maerkisch19}, which we recommend if scalar and tensor interactions are assumed to be absent.
The $90$\% confidence region corresponding to the Fierz result in Eq.~\eqref{eqn:final-result} is
\begin{equation}
  -0.018 \leq b \leq 0.052.
\end{equation}
To date this is the most precise limit on $b$ obtained from a single measurement in neutron decay.

%% TODO:  statement on blinding. Which result to choose for BSM or SM applications 

%\section{\label{sec:summary}Summary}

In conclusion, we have demonstrated the extraction of limits on the Fierz interference term from a measurement of beta asymmetry spectra in \emph{polarized} neutron decay.
In contrast to a more direct determination from the spectrum in \emph{unpolarized} neutron decay, this method profits from much better controlled systematics and is limited by available statistics.
Next generation instruments like PERC \cite{perc08, perc12} will allow the measurement of decay correlations with strongly improved statistical uncertainties.
The correlated analysis of $A$ and $b$ in these measurements will become even more important and serve as valuable input to derive limits on scalar and tensor couplings from beta decay data \cite{Gonzalez-Alonso18}.

%\section{\label{sec:ack}Acknowledgements}

\begin{acknowledgments}
The authors acknowledge the excellent support of the services of the Physikalisches Institut, Heidelberg University, and the ILL.  We thank D.~Rechten, TUM, for providing us with ultra-thin carbon foils, and U.~Schmidt, Heidelberg, and M.~González-Alonso, València, for valuable discussions. We thank D.~Dubbers for his support and enthusiasm.
% Funding
This work was supported by the Priority Programme SPP~1491 of the German Research Foundation, contract nos. MA~4944/1-2, AB~128/5-2 and SO~1058/1-1, the Austrian Science Fund (FWF) contract nos. P~26636-N20, and W1252-N27 (DK-PI), the German Federal Ministry for Research and Education, contract nos. 06HD153I and 06HD187, and the DFG cluster of excellence 'Origin and Structure of the Universe'. The computational results presented have been achieved in part using the Vienna Scientific Cluster (VSC).
\\Note added: Ref.~\cite{Sun20} presents a similar analysis by the UCNA collaboration and extends the analysis of the beta spectrum from unpolarized neutron decay of \cite{Hickerson17}.
\end{acknowledgments}

%% REFERENCES
%\bibliography{bibliography/beta_spectroscopy,%
%  bibliography/correlation,%
%  bibliography/general,%
%  bibliography/neutron_background,%
%  bibliography/neutron_decay_corrections,%
%  bibliography/neutron_instrumentation,%
%  bibliography/programming,%
%  bibliography/thesis,%
%  bibliography/other_measurements}

\end{document}